\documentstyle[12pt]{article}
\begin{document}
\begin{titlepage}
\baselineskip 0.35cm
\begin{center}
\hfill SNUTP 93-85\\
\null\hfill \\
\vskip 1.5cm
{\large \bf  Small SUSY phases in string-inspired supergravity}
\vskip 2.5cm
{Kiwoon Choi}
\vskip .05cm
{ Department of Physics, Chonbug National University}
\vskip 0.05cm
{Chonju, 560-756 Korea}

\vskip 2.5cm

{\bf Abstract}
\end{center}
\begin{quotation}
\baselineskip .73cm
In supersymmetric models, there are new CP violating phases which,
if unsuppressed, would give a too large neutron electric dipole moment.
We examine the possibility of small SUSY phases
in string-inspired supergravity models
in which supersymmetry is broken
by the auxiliary components of the dilaton and moduli superfields.
It is found that the SUSY phases can be suppressed
by a small factor governing the breakdown
of the approximate Peccei Quinn symmetries
nonlinearly realized for the moduli superfields that
participate in supersymmetry breaking.
In many cases, the symmetry breaking factors are exponentially
small for moderately large values of the moduli, leading
to small phase  values in a natural way.

\end{quotation}
\end{titlepage}
\baselineskip .75cm

\eject

It is well known that  supersymmetric models  have
new sources of CP violation other than the QCD angle and the
Kobayashi-Maskawa phase
that exist already in the standard model.
If these new CP-violating phases are unsuppressed and also
the superpartners have masses around 100 GeV,  the resulting
neutron electric dipole moment would  exceed the current experimental
bound by  a factor of $10^{2}-10^{3}$ $\cite{susyedm}$.
To resolve this difficulty,
one needs to have either the new phases smaller than $10^{-2}-10^{-3}$
or superpartners, particularly the squarks,
having masses  greater than a few TeV $\cite{garisto}$.
Although the option of heavy superpartners is still possible,
it is more customary to take the small phase option
while keeping the superpartner masses
to be around 100 GeV.
Then the required smallness is somewhat disturbing,
and thus it is desirable to have any explanation
for the small phase values.

The above small phase problem is in fact a problem
of supersymmetry (SUSY) breaking since the phases
and also  the superpartner masses  are determined mainly by
the parameters of soft SUSY breaking terms.
Presently the most popular way to break SUSY
is to utilize  a hidden sector
in the context of $N=1$ supergravity theories $\cite{nilles1}$.
Among phenomenologically acceptable supergravity
models, those from superstrings are
of particular interests.
In string-inspired supergravity models,
it is commonly assumed that
supersymmetry is broken
by the auxiliary components of the dilaton and/or moduli
superfields whose superpotential is induced
by nonperturbative hidden sector dynamics $\cite{nilles2}$.
In this paper, we explore the possibility of small
SUSY phases in string-inspired supergravity models
adopting this scenario of SUSY breaking $\cite{ibanez2}$.
It is found that the SUSY phases can be suppressed
by a small  factor governing the breakdown of the
approximate Peccei Quinn (PQ)
symmetries that are associated with the pseudoscalar
components of the  moduli $\cite{witten}$.
The novel feature of this suppression mechanism is that it is
completely independent of how CP is broken $\cite{barr}$.
If the PQ symmetries for the moduli that participate in
SUSY breaking are good enough, then the resulting
SUSY phases would be small enough.
However if a modulus whose PQ symmetry
is badly broken contributes to SUSY breaking significantly,
it looks quite difficult to achieve
the small SUSY phases unless a strong assumption
on CP violation is made.


To begin with, let us define the SUSY phase problem
more precisely. We consider a low energy supersymmetric
model  with the superpotential
\begin{equation}
W=\lambda_{ijk} \Phi_i\Phi_j\Phi_k+\mu H_1H_2,
\end{equation}
and the soft breaking part containing
\begin{equation}
\frac{1}{2}m_a\lambda_a\lambda_a+A_{ijk}\varphi_i\varphi_j\varphi_k
+B h_1h_2,
\end{equation}
where $\Phi_i$ denote generic  chiral  superfields
with their scalar components
$\varphi_i$,  $\lambda_a$ are gauginos, and $H_{1,2}$
($h_{1,2}$) stand for the two Higgs superfields (their scalar
components) in the minimal supersymmetric standard model.
If one does not make any assumption
on CP, all parameters above are complex in general.
Then compared to the non-supersymmetric counterpart,
the theory  contains new CP-violating phases:
\begin{equation}
\phi_A=\{{\rm arg}(\frac{A_{ijk}}{\lambda_{ijk}})\}, \, \,
\phi_B=\{{\rm arg}(\frac{B}{\mu})\}, \, \,
\phi_C=\{{\rm arg}(m_a)\}.
\end{equation}
These phases, more precisely the combinations
$\phi=\{\phi_A-\phi_C, \phi_B-\phi_C\}$,  give the neutron
electric dipole moment
$d_n\simeq (10^{-22}-10^{-23})\times \sin \phi$ e-cm
for the superpartner masses around 100 GeV $\cite{susyedm}$.
Then the phases are constrained as $\phi\leq 10^{-2}-10^{-3}$.
Of course, as we have  mentioned, one can relax this constraint
by assuming that the squark masses are larger.

Let us consider a supergravity  model which is assumed
to enjoy the following  properties of superstring vacua.
First of all, the model contains
a hidden sector which  generally has a large gauge
group as well as matter fields
that transform nontrivially under the hidden sector gauge group.
Also the model contains
the dilaton  multiplet $S$ and the moduli multiplets
which describe a variety of deformations of the internal
space. Among the moduli,
we consider only  the overall modulus $T$ for a moment.
Later we will discuss the
effects of including other moduli.
The dilaton component ${\rm Re}(S)$
couples to the gauge kinetic terms, giving
the gauge coupling constant as $g^2=1/{\rm Re}(S)$
at string tree level.
The modulus field ${\rm Re}(T)$
characterizes the size of the internal space.
Then $1/{\rm Re}(T)$ corresponds to the sigma model coupling constant.
The imaginary components ${\rm Im}(S)$ and ${\rm Im}(T)$
correspond to the model-independent axion and the
internal axion respectively.

As is well known, a four-dimensional $N=1$ supergravity
action is characterized by the K\"{a}hler potential $K$,
the superpotential $W$,  and the gauge kinetic function $f_a$
for the $a$-th gauge group.
At the compactification scale,
one may expand the K\"{a}hler and superpotential in
chiral matter fields $\Phi_i$ as
\begin{eqnarray}
&&K=\tilde{K}+Z_{ij}\Phi_i\bar{\Phi}_j+(Y H_1H_2
+{\rm h.c.})+..., \nonumber \\
&&W=\tilde{W}+\tilde{\mu}H_1H_2
+\tilde{\lambda}_{ijk}\Phi_i\Phi_j\Phi_k+...,
\end{eqnarray}
where all coefficients in the expansion
are generic functions
of  $S$ and $T$,
and the ellipses stands for higher order terms.
In perturbation theory, $\tilde{W}$ and $\tilde{\mu}$ vanish,
but they are induced by nonperturbative effects
in the hidden sector which is integrated out already.
Although the wavefunction factor $Z_{ij}$
can have an off-diagonal element in general,
here we assume it is diagonal, viz $Z_{ij}=Z_i\delta_{ij}$,
for the sake of simplicity.
At any rate, off-diagonal elements are required to be small,
roughly smaller than   $10^{-2}Z_i$,
to avoid a too large flavor changing neutral current effect.
In string theory,  CP corresponds  to  a
discrete  gauge symmetry $\cite{gaugecp}$,
and thus it must be broken spontaneously.
However if broken  at the compactification scale,
it would appear to be explicitly broken
in the $d=4$ effective lagrangian.
Here we do not make any assumption on the
nature of CP violation, and thus
allow all complex parameters in the K\"{a}hler and superpotential
to have the phases of order unity in general $\cite{foot3}$.

Let us now assume that SUSY is broken
by the auxiliary components of the dilaton $S$ and
the overall modulus $T$:
\begin{equation}
\bar{F}_I=e^{\tilde{K}/2}|\tilde{W}|(\bar{\partial}_I\partial_J\tilde{K})^{-1}(\partial_J
\tilde{K}+\partial_J\ln \tilde{W}),
\end{equation}
where the indices $I,J=S,T$.
It is then straightforward to derive the
resulting global SUSY theory
together with the soft breaking terms $\cite{ibanez2,louis}$.
If one writes the effective superpotential (for un-normalized fields) as
in eq. (1),
\begin{equation}
\lambda_{ijk}=e^{-i\xi}e^{\tilde{K}/2}\tilde{\lambda}_{ijk}, \quad
\mu=\mu_1+\mu_2+\mu_3,
\end{equation}
where
$\mu_1=\lambda\langle{N}\rangle$,
$\mu_2=(m_{3/2}-\bar{F}_I\bar{\partial}_I)Y$,
$\mu_3=e^{-i\xi}e^{\tilde{K}/2}\tilde{\mu}$ ($\xi={\rm arg}(\tilde{W})$),
and  $m_{3/2}=e^{\tilde{K}/2}|\tilde{W}|$
denotes the gravitino mass. Note that here we consider three
possible sources of the $\mu$-term. Amongst them,
the $\mu_1$-piece
is obtained by replacing the singlet field $N$
which has the Yukawa coupling $\lambda NH_1H_2$ by its vacuum value.
If we do not have any
such singlet as in the minimal supersymmetric standard model,
then $\mu_1=0$.
For the soft terms written as in eq. (2), one finds $\cite{louis}$
\begin{eqnarray}
&&m_a=\frac{1}{2}g_a^2F_I\partial_If_a, \nonumber \\
&&A_{ijk}=\lambda_{ijk}F_I\partial_I[\ln(e^{\tilde{K}}\tilde{\lambda}_{ijk}/Z_iZ_jZ_k)],
\nonumber \\
&& B=B_1+B_2+B_3,
\end{eqnarray}
where
\begin{eqnarray}
&&B_1/\mu_1=A_{\lambda}/\lambda, \nonumber \\
&&B_2/\mu_2=F_I\partial_I[\ln(e^{\tilde{K}/2}\mu_2/Z_{H_1}Z_{H_2})]-m_{3/2},
\nonumber \\
&&B_3/\mu_3=F_I\partial_I[\ln(e^{\tilde{K}}\tilde{\mu}
/Z_{H_1}Z_{H_2})]-m_{3/2}.
\end{eqnarray}
Here $A_{\lambda}$ denote the $A$-coefficient of the trilinear
soft term for the term $\lambda NH_1H_2$ in the superpotential.
These  soft parameters are defined  at the compactification
scale while the experimental constraints stand for those
defined at the weak scale.
For the phases $\phi_A={\rm arg}(A_{ijk}/\lambda_{ijk})$,
$\phi_B={\rm arg}(B/\mu)$, and $\phi_C={\rm arg}(m_a)$, this
point is not so relevant since
$\phi_{A,B,C}$ at the weak scale remain to be small
enough as long as
they are less than $10^{-2}-10^{-3}$ at the compactification scale
$\cite{rgrunning}$.

The above formulae for  soft terms show that there are  a lot
of potentially complex quantities which can contribute
to the phases $\phi_{A,B,C}$.
First of all, the SUSY breaking order parameters $F_I$
can be complex in general. A nonzero ${\rm arg}(F_I)$ may arise
due to  nonzero vacuum values of ${\rm Im}(S)$ and ${\rm Im}(T)$,
or due to the complex Yukawa couplings of hidden matters which
would affect
the induced superpotential $\tilde{W}$.
Furthermore, although $\tilde{K}$ and $Z_i$ are real functions,
their derivatives $\partial_I{\tilde{K}}$,
$\partial_I\bar{\partial}_J\tilde{K}$, and $\partial_IZ_i$
can be complex.  Besides these, we can have
complex $\partial_If_a$, $\partial_I\ln(\tilde{\lambda}_{ijk})$, $\partial_I\ln
(Y)$,
$\partial_I\ln(\tilde{\mu})$, and several others.
It is then  convenient
to  classify all the relevant (potentially)
complex quantities as follows:
\begin{eqnarray}
&&X_1: \partial_I\tilde{K}, \partial_I\bar{\partial}_J\tilde{K}, \partial_I
Z_i, \partial_I f_a,
\partial_I\ln(\tilde{\lambda}_{ijk}); \nonumber  \\
&&X_2: \partial_I\ln (\tilde{W}); \quad X_3:\partial_I\ln (\tilde{\mu});
 \nonumber \\
&&X_4: \partial_I\ln (Y), \bar{\partial}_I\ln (Y),
\partial_I\bar{\partial}_J\ln (Y).
\end{eqnarray}
It is easy to see that if  $X_1$ and $X_2$ are all real,
then $\phi_A$ and $\phi_C$ do vanish.
The phase $\phi_B$ is affected also by $X_3$ and $X_4$,
and thus making it small requires more conditions.

At first sight,
it looks very nontrivial to make all the above quantites
real. However as we will see, due to
the approximate  PQ symmetries nonlinearly realized
for ${\rm Im}(S)$ and ${\rm Im}(T)$,
many of them are in fact (approximately) real.
In spacetime and world sheet perturbation theory,
the vertex operators of
${\rm Im}(S)$ and ${\rm Im}(T)$ vanish at zero momentum $\cite{witten}$.
Then the corresponding
perturbative effective action would be invariant under the
PQ symmetries:
\begin{equation}
U(1)_S: S\rightarrow  S+i\alpha_S, \quad U(1)_T: T\rightarrow T+i\alpha_T,
\end{equation}
where $\alpha_{S,T}$ are arbitrary real constants.
The symmetry $U(1)_T$ is expected to be broken
by nonperturbative effects on world sheet, i.e. world sheet
instantons $\cite{dine}$,
even at string tree level while $U(1)_S$ is broken only
by nonperturbative effects on spacetime.
As a result, their breakdown is suppressed
either by $e^{-c_1 S}$ (for $U(1)_S$) or by $e^{-c_2 T}$
(for $U(1)_T$) where $c_1$ and $c_2$ are some real constants.
For $S$ normalized as $g^2=1/{\rm Re}(S)$ at string tree level,
$c_1$ is of $O(4\pi^2)$.
Then for the phenomenologically favored $g^2\simeq 1$,
we have $|e^{-c_1 S}|\ll 10^{-3}$.
This implies that
one can safely ignore $U(1)_S$-violating effects
for the discussion of the SUSY phases
if they are  {\it not} the leading effects, but
just give small corrections to the leading perturbative effects.
A common normalization of $T$ is
that of $T\equiv T+i$  for which  the world sheet
instanton factor is  given by
$q\equiv e^{-2\pi T}$.
Then  all $U(1)_T$-violating corrections are suppressed
by a factor of $O(q)$.

For the supergravity action invariant
under  $U(1)_S$ and $U(1)_T$,
the corresponding K\"{a}hler potential can be chosen to be invariant.
Then the superpotential should be invariant up to
a constant phase and the gauge kinetic functions
up to  imaginary constants $\cite{burgess}$. This  implies
that (i) $\tilde{K}$ and $Z_i$ are the real functions
of the real variables ${\rm Re}(S)$ and ${\rm Re}(T)$
 up to corrections of $O(q)$,
(ii) the gauge kinetic functions
$f_a=\tilde{k}_aS+\tilde{l}_aT+O(q)$ with some
real constants  $\tilde{k}_a$
and $\tilde{l}_a$, (iii) the Yukawa couplings
$\tilde{\lambda}=(1+O(q))\tilde{\lambda}_0\exp (aS+bT)$
where  $a$ and $b$ are some real constants
while the constant $\tilde{\lambda}_0$ is  complex in general.
In (iii), the constants $a$ and $b$ also can carry
the omitted flavor indices $\cite{foot1}$.
Clearly  the  properties (i), (ii), (iii)
ensure that the quantities
of $X_1$ are all real up to corrections of $O(q)$.

A complex  $\partial_I\ln (\tilde{W})$ would give a nonzero
phase of the SUSY breaking order parameters $F_I$.
In string theory, $\tilde{W}$ is induced by
nonperturbative  hidden sector dynamics.
All renormalizable interactions of the hidden sector
fields are determined by the
hidden gauge kinetic functions $f_h$
and the hidden Yukawa couplings $\tilde{\lambda}_h$
(up to the scaling due to non-minimal K\"{a}hler potential).
Then $\tilde{W}$ would appear as
a generic holomorphic function of
the holomorphic quantities $f_h$ and $\tilde{\lambda}_h$.
Since it is nonperturbative in the hidden gauge coupling
constant $1/{\rm Re}(f_h)$, $\tilde{W}$
is suppressed by some powers of $e^{-f_h}$.
The forms of $f_h$ and $\tilde{\lambda}_h$ are restricted
by $U(1)_S$ and $U(1)_T$ as in
(i) and (ii). Then using the arguments involving
anomaly free $R$-symmetries and dimensional analysis $\cite{amati}$,
$\tilde{W}$ can be written as $\cite{foot2}$:
\begin{equation}
\tilde{W}=\sum_{n=1}^{N_W} W_n=\sum d_n(T) \exp (k_{n}S+l_{n}T)
\end{equation}
where $k_{n}$ and $l_n$ are some {\it real} constants
and $d_{n}(T)=\hat{d}_n(1+O(q))$ for a complex constant $\hat{d}_n$.

Since the corrections less then $10^{-2}-10^{-3}$
are essentially ignored
in our approximation, $\tilde{W}$
includes only the terms such that $|W_n/W_1|\geq 10^{-2}-10^{-3}$
where $W_1$ denotes the term with the largest vacuum value.
Clearly  the number of such terms, viz $N_W$, would depend on the
details of the hidden sector, e.g. on the number
of simple hidden gauge groups, the ratios of the dynamical mass scales,
and also the Yukawa couplings.
Let us briefly discuss the number of terms in $\tilde{W}$
for several simple cases.
If the hidden sector contains a simple gauge group ${\cal G}_1$
whose  dynamical mass scale $\Lambda_1$ is far above those of
other groups, then  $\tilde{W}\simeq W_1\sim \Lambda_1^3$
where $W_1$ contains the  gaugino condensation
together with possible matter condensations.
In the case that there exists  another simple
group ${\cal G}_2$ with $\Lambda_2$ comparable to $\Lambda_1$,
$\tilde{W}$ contains  at least two terms
$W_{1,2}\sim \Lambda_{1,2}^3$.
If the gaugino condensations
are largely dominate over other possible contributions,
e.g.  matter condensations, one simply has $N_W=2$
associated with the two gaugino condensations of ${\cal G}_1$ and ${\cal G}_2$.
Even in the case that matter condensations become important,
if the fields that transform nontrivially
under ${\cal G}_1$ communicate weakly with those
of  ${\cal G}_2$, e.g.
communicate only via nonrenormalizable interactions,
one still has $N_W=2$ but now
$W_1$ and $W_2$ contain both the gaugino and matter condensations
of the ${\cal G}_1$-sector and the ${\cal G}_2$-sector respectively.

As is well known, the case of $N_W=1$ suffers from the runaway
of the dilaton. Thus let us consider the
next simple case of $N_W=2$ which has been argued to be able
to produce  phemenologically interesting results $\cite{casas}$.
In fact, most of $\tilde{W}$'s analyized in the literatures
have $N_W=2$.
It is  obvious that  $\partial_I\ln(W_{1,2})$ is real up to
corrections of $O(q)$.
However to have  a real $X_2=\partial_I\ln(\tilde{W})$,
one still needs the
relative phase ${\rm arg}(W_2/W_1)$ to be  CP conserving.
Interestingly enough, this can be achieved dynamically
by the vacuum value of  the model-independent
axion ${\rm Im}(S)$.
Using the standard scalar potential in supergravity,
one easily find the following
form of the axion-potential:
\begin{equation}
V_{\rm axion}=\Omega \,  [\cos ({\rm arg}({W}_2/{W}_1))+O(q)],
\end{equation}
where ${\arg}({W}_2/{W}_1)=(k_2-k_1){\rm Im}(S)+\delta$, and
$\Omega$ and $\delta$
are  real functions which are {\it independent} of ${\rm Im}(S)$.
Clearly minimizing this axion potential leads to
a real  value of $W_2/W_1$ up to $O(q)$,
and thus a real value of  $\partial_I\ln (\tilde{W})$ up to
$O(q)$.
Note that  here $\delta={\rm arg}(d_2/d_1)+(l_2-l_1){\rm Im}(T)$
is of order unity in general, but
it is dynamically relaxed to a CP conserving value
by the vacuum value of ${\rm Im}(S)$. This is quite similar to the
Peccei-Quinn mechanism $\cite{pq}$ relaxing $\theta$
in the axion solution to the strong CP problem.
At any rate, now we find that  $X_1$ and $X_2$ are real up to
corrections of $O(q)$, and thus $\phi_{A,C}=O(q)$
if   SUSY breaking is due to
the auxiliary components of $S$ and $T$ with $N_W= 2$.

In the above, we have noted  the dynamical relaxation
of the relative phases between $W_n$'s  for $N_W=2$.
This mechanism can
be easily generalized for  $N_W>2$.
Suppose we have $N_A$ axion-like fields $\vec{A}=
(A_1, ..., A_{N_A})$ whose nonderivative couplings
are {\it mainly} given by $W_n\sim e^{i \vec{c}_n\cdot\vec{A}}$
due to the associated nonlinear
PQ symmetries: $\vec{A}\rightarrow \vec{A}+\vec{\alpha}$.
Here $\vec{c}_n$ and $\vec{\alpha}$ denote some real constant vectors.
Note that we always have two such fields, ${\rm Im}(S)$ and
${\rm Im}(T)$, and thus $N_A\geq 2$. With more moduli,
$N_A$ would become larger. Let $N$ denote the number
of linearly independent vectors among $\{\vec{c}_n-\vec{c}_m\}$.
Then for  $N_W\leq N+1$,
all the relative phases ${\rm arg}(W_n/W_m)$
are relaxed to CP conserving values
(of course up to corrections associated with the breakdown
of the PQ symmetries).
Note that  $1\leq N\leq N_A$ in general, but it is
quite conceivable to have $N=N_A$.

The discussion of the remained phase $\phi_B$
is more model-dependent since
presently there is no definite theory for the $\mu$-term $\cite{mu}$.
Here we consider three simple scenarios in which one of
$\mu_{1,2,3}$  dominates over the other two
by more than a factor
of $10^2$ to $10^{3}$. In the first case that
$\mu_1=\lambda\langle N\rangle$ dominates,
$\phi_B$ simply corresponds to $\phi_A$ and thus is of $O(q)$.
In the second case that $\mu_2=(m_{3/2}-\bar{F}_I\bar{\partial}_I)Y$ dominates,
$\phi_B$ would
receive additional contribution from $X_4$.
Orbifold compactifications give $Y=0$ $\cite{louis}$ and thus  they
do not correspond to this case.
It has been pointed out that
for $(2,2)$ Calabi-Yao compactifications
$Y$ is related to some Yukawa couplings $\cite{louis}$
by world sheet Ward identities.
One then has $X_4=O(q)$ $\cite{ibanez2}$, implying
$\phi_B=O(q)$.
In the third case that $\mu_3$ dominates, $\phi_B$ would receive
a contribution from $X_3=\partial_I\ln (\tilde{\mu})$.
Since $\tilde{\mu}$ is  due to nonperturbative effects,
it can be written as $\tilde{\mu}=\sum \tilde{\mu}_n$
where  $\tilde{\mu}_n=(1+O(q))z_n\exp (x_n S+y_n T)$.
Here $x_n$ and $y_n$ are some real constants
while $z_n$ is a complex constant.
Again  $\partial_I\ln(\tilde{\mu}_n)$ is real up to $O(q)$,
however to have real $\partial_I\ln(\tilde{\mu})$
one still needs the relative phases of $\tilde{\mu}_n$
to be CP conserving.
For the induced superpotential $\tilde{W}$,
the relative phases of $W_n$
could be relaxed to CP conserving values by the vacuum
values of the axion-like fields.
For $\tilde{\mu}$, we do not have any such mechanism.
Of course, if there is just a single term in $\tilde{\mu}$,
$\partial_I\ln(\tilde{\mu})$ is real up to $O(q)$
and thus $\phi_B=O(q)$.
If there are more than one term, e.g. $\tilde{\mu}_{1,2}$ such that
$\tilde{\mu}_1>\tilde{\mu}_2$, both ${\rm arg}(\partial_I\ln(\tilde{\mu}))$
and $\phi_B$  would be of $O(\tilde{\mu}_2/\tilde{\mu}_1)$
in general.

So far, our discussion has been restricted
to the case with $S$ and $T$ only.
It is in fact straightforward to extend
the analysis to the case with more moduli.
Let us suppose an additional modulus $M$
and define the corresponding PQ symmetry
$U(1)_M: M\rightarrow M+i\alpha_M$.
This additional modulus can affect our
previous analysis by two ways.
First of all, it can directly
affect the SUSY phases by participating
in SUSY breaking, i.e.  by having a nonzero
auxiliary component $F_M$.
Secondly it can affect
$F_I$ ($I=S,T$) via the wave function mixing with $S$ and $T$.
Let $q_M$ denote a factor characterizing the size of $U(1)_M$-breaking
corrections.
Then including  $M$ in the analysis,
it is easy to see that
the SUSY phases receive additional contributions
which are  of the order of
either $q_MF_M/m_{3/2}$ or
$q_M\bar{\partial}_I\partial_M\tilde{K}/\bar{\partial}_I\partial_I\tilde{K}$.
Thus even in the case with more moduli,
the SUSY phases are  suppressed by a factor governing
the breakdown of the PQ symmetries nonlinearly realized
for the moduli that participate in SUSY breaking.

What would be the typical size of the PQ symmetry
breakings?
For a K\"{a}hler class modulus $M_K$ that is associated
with
the deformation of the K\"{a}hler class of the internal space,
e.g. the overall modulus $T$,
the pseudoscalar component ${\rm Im}(M_K)$ comes from
the zero modes of the antisymmetric tensor field.
Then the corresponding PQ symmetry is
broken only by world sheet instantons $\cite{witten}$,
leading to
$q_{M_K}=e^{-2\pi M_K}$ which can be small enough
to give the SUSY phases less than $10^{-2}-10^{-3}$
for a moderately large value of ${\rm Re}(M_K)$.
For another type of moduli, the complex structure
moduli $M_C$ that is associated with the deformation
of the complex structure, the size of $q_{M_C}$ is somewhat
model-dependent.
For orbifold compactifications, $q_{M_C}$ is exponentially
small due to the modular symmetry $SL(2,Z)$ $\cite{ibanez1}$.
However for Calabi-Yao cases, $q_{M_C}$ can be  of order
unity even at the leading order approximation $\cite{witten}$.
As a result, to achieve small SUSY phases in Calabi-Yao compactification,
one needs to assume that the complex structure moduli
give negligible contribution to SUSY breaking,
viz $F_{M_C}/m_{3/2}\leq 10^{-2}-10^{-3}$, and also have
small wave function mixings with the K\"{a}hler moduli $M_K$,
viz $\bar{\partial}_{M_C}\partial_{M_K} \tilde{K}
/\bar{\partial}_{M_K}\partial_{M_K}\tilde{K}\leq
10^{-2}-10^{-3}$, or needs some assumption on CP violation.

Barring the dynamical relaxation
of the relative phases in $\tilde{W}$,
it is a general conclusion in string-inspired
supergravity that the SUSY phases  $\phi_{A,C}$ are suppressed
by a factor governing the breakdown of the PQ symmetries
nonlinearly realized for the moduli that participate in SUSY breaking.
A similar suppression can occur also for $\phi_B$
if the $\mu$-term is dominated by  either $\mu_1$ or $\mu_2$.
In many cases, particularly for generic K\"{a}hler
class moduli,  the PQ symmetry breaking factors
are exponentially small for a moderately
large values of the moduli.
Then one would achieve small SUSY phases,
i.e.  $\phi_{A,B,C}\leq 10^{-2}-10^{-3}$,
in a quite natural way in string-inspired supergravity
models.
In regard to the dynamical relaxation mechanism,
we have argued that it can take place for $N_W\leq N_A+1$
where $N_A$ denotes the number of the axion-like fields which is
essentially the same as the number
of moduli (including $S$) with good PQ symmetries.
Most of $\tilde{W}$'s analyized in the literatures
have $N_W=2$ for which the relaxation always occurs.



\begin{thebibliography}{99}
\bibitem{foot3}
In some cases, complex parameters in the supergravity
action get nonzero phases due to the complex vacuum values of some
moduli other than $T$, e.g. the complex structure moduli in Calabi-Yao
compactifications. At this stage,
it is assumed that such moduli are frozen
to their vacuum values without breaking supersymmetry.
We will discuss more general cases later.
\bibitem{foot1}
In fact, one can always set $a=b=0$ via a
holomorphic field redefinition: $\Phi_i\rightarrow \gamma_i\Phi_i$
where $\gamma_i=(1+O(q))\exp (\alpha_i S-\beta_i T)$ with
real constants $\alpha_i$ and $\beta_i$.
This would  lead to the changes of $Z_i$ and $f_a$ also,
but the properties (i) and (ii) are not affected.
In cases with $SL(2,Z)$ modular invariance, e.g. orbifold
models,  a proper choice
of $\gamma_i$ would be $\gamma_i=e^{\alpha_iS}[\eta (T)]^{12\beta_i/\pi}$
where $\eta(T)$ denotes the Dedekind eta function.
\bibitem{foot2}
In some cases, $\tilde{W}$ can be induced by world sheet
instantons $\cite{dine}$, but it can still be written as eq. (11).
\end{thebibliography}
\end{document}